\def\arxivpreprint{}

\documentclass{article} 

\usepackage[T1]{fontenc}

\usepackage[dvipsnames,svgnames,table]{xcolor}
\ifdefined\arxivpreprint
  \usepackage[preprint]{iclr2027_conference}
\else
  \usepackage{iclr2027_conference}
\fi
\usepackage{times}

\usepackage{microtype}

\usepackage{graphicx}
\graphicspath{{figs/final/}{../../exps/2026-0831-concealment-phase-transition/notes/figs/}}
\usepackage{booktabs}
\usepackage{fvextra}  %

\usepackage{amsmath}
\usepackage{amssymb}

\IfFileExists{dsfont.sty}
  {\usepackage{dsfont}}
  {\IfFileExists{bbm.sty}
    {\usepackage{bbm}}
    {}}

\newcommand{\secref}[1]{Section~\ref{sec:#1}}

\newcommand{\appref}[1]{Appendix~\ref{sec:#1}}

\definecolor{light-gray}{gray}{0.8}

\usepackage[colorlinks,allcolors=NavyBlue]{hyperref}
\usepackage{url}

\title{Multi-agent discussion gains less when dissent is withheld}

\author{%
Chand Sahil Mansuri$^{1}$\quad Xin (Vision) Wang$^{1}$\quad Mengying Li$^{2}$\quad Bryan Acton$^{2}$ \\
\textbf{Rory Eckardt$^{2}$\quad Dhaval Patel$^{3}$\quad Sadamori Kojaku$^{1}$\thanks{Corresponding author: \texttt{skojaku@binghamton.edu}}} \\[4pt]
$^{1}$School of Systems Science and Industrial Engineering,\quad $^{2}$School of Management \\
Binghamton University, Binghamton, NY, USA \\
$^{3}$IBM T.~J.\ Watson Research Center, Yorktown Heights, NY, USA
}

\begin{document}

\maketitle

\begin{abstract}

Multi-agent systems of LLMs add discussion to majority voting and are therefore expected to be more capable. However, empirical reports conflict on whether discussion improves accuracy or leads to an incorrect consensus. Here, we introduce a parsimonious model that explains when discussion improves accuracy and when it ends in an incorrect consensus, built from four behaviors repeatedly observed in LLM agents: (1) withholding dissent, (2) internalizing a stated answer, (3) reconsidering after seeing dissent, and (4) correcting toward the correct answer. The model shows that discussion can overturn an incorrect initial majority only when the withholding rate $c$ is below a critical rate $c^* = \gamma/(\gamma + a)$, set by the net correction rate $\gamma$ and the internalization rate $a$. We estimate these rates from conversation logs with a Bayesian method and place LLM teams relative to $c^*$.
As the model predicts, the gain from discussion shrinks as withholding rises, across LLMs and on a hidden profile benchmark, HiddenBench, and MedEInst.
Instructing agents not to withhold dissent increases this gain.
Turning reasoning off also increases the gain, because reasoning raises the internalization rate $a$ and keeps agents from reconsidering a minority answer.
These findings reconcile the conflicting reports and identify when discussion outperforms majority voting.

\end{abstract}

\section{Introduction}\label{sec:1}

Does discussion make a group more accurate than majority voting over the independent answers of its members?
Galton's argument on the wisdom of crowds explains why majority voting performs surprisingly well: nearly 800 individuals at a livestock exhibition estimated an ox's weight largely inaccurately, but their median closely matched the true weight~\citep{galton1907vox}, because independent errors cancel out~\citep{grofman1983thirteen, wang2023selfconsistency}.
However, Galton's argument hinges on an important premise, namely that the votes are independent of one another.
This premise does not hold when people or large language model (LLM) agents reach a consensus, because they can discuss and influence one another's answers.
As members see one another's views, the majority exerts pressure to conform~\citep{asch1956studies}, and the group converges on a shared answer without becoming more accurate~\citep{lorenz2011wisdom}.
At the same time, discussion can rescue the group from an incorrect majority: members who are right can convince the rest even when they are in the minority.
Whether discussion outperforms majority voting therefore depends on whether members voice their dissent or withhold it.

LLM multi-agent systems can fall short~\citep{weng2025conformity, zhu2025conformity, zhang2024exploring, sharma2024towards}.
If conformity were the only effect, discussion would make the group less accurate, but the empirical reports are inconsistent~\citep{choi2025debate, estornell2024debate, huang2024selfcorrect, wang2024token, zhang2025stop}.
Some studies show that discussion, often run as multi-agent debate, outperforms standalone reasoning or voting~\citep{du2024improving, liang2024encouraging, chen2024reconcile}, whereas others report no effect~\citep{smit2024mad, wang2024rethinking, kaesberg2025voting}.
The same conformity phenomenon is shown to benefit or harm performance depending on the LLM and task setting~\citep{zhang2024exploring}.
Yet it remains unclear what factors produce these divergent consensus outcomes.

We propose a simple mathematical model that explains the success and failure of conforming teams from two quantities, the withholding rate and the single-agent accuracy, the probability that an agent answers correctly without discussion.
Together, the two quantities separate cases where discussion outperforms majority voting from cases where the team ends in an incorrect consensus.

While prior studies of LLM teams diverge on which mechanisms drive the outcome of discussion, we focus on the intersection of their observations, not their union, to provide a parsimonious model that applies across multi-agent systems.
This intersection consists of four behaviors that these studies repeatedly report.
The first behavior is withholding, in which agents privately dissent while publicly aligning with the majority~\citep{weng2025conformity, ys2026pluralistic, pandey2026knowwrong}.
The second is internalization, in which agents come to adopt the answer they stated as their own private belief~\citep{weng2025conformity, hao2026flips}.
The third is that visible dissent from others prompts agents to reconsider~\citep{stechly2025selfcritique, jiang2024selfdiscrimination}.
The fourth is that this reconsideration tends to reach the correct answer.
Withholding suppresses visible dissent, removing the triggers needed for correction.
As internalization turns conformed statements into private beliefs, even teams starting with a correct majority can end in an incorrect consensus~\citep{wang2024rethinking, wu2025debate, okawa2026biased}.

From the minimal specification of these four behaviors, we derive a critical withholding rate $c^*$.
Below $c^*$, discussion can outperform majority voting even when starting from an incorrect majority.
Conflicting empirical reports on discussion efficacy then correspond to different regimes on one phase diagram over the single-agent accuracy $p$ and the withholding rate $c$, on which each region is labelled by where the team ends up.

To test these predictions empirically, we must determine whether real LLM teams operate in the recovery regime, where the initial majority is wrong ($p < 1/2$) yet discussion still reaches the correct answer, or end in an incorrect consensus.
We develop a Bayesian method to estimate the model parameters directly from multi-agent conversation logs.
Across settings, instructions systematically shift how often agents withhold their opinions. 
Specifically, instructions encouraging honesty reduce withholding, whereas instructions valuing cohesion push teams farther above the threshold.
As teams move further above the critical threshold, the benefit they gain from discussion compared to majority voting diminishes.
This relationship between the distance from the threshold and discussion gain holds in different benchmarks, HiddenBench, MedEInst, and MuSiQue. 

Our model explains two pathways by which an LLM team can arrive at an incorrect consensus: one is that agents who disagree remain silent (a high withholding rate $c$), so the team never hears the dissent it needs to correct itself; the other is that dissent is stated but fails to shift the team, because agents adhere to what they have already stated (a high internalization rate $a$) and rarely reach the correct answer when they reconsider (a low net correction rate $\gamma$), which together lower the critical threshold $c^*$.
Instructions remedy the first pathway by reducing the withholding rate $c$.
However, if the second pathway is open, voicing dissent is insufficient, because agents with strong internalization do not change their answers when a minority disagrees with them.
Reasoning models exemplify this situation: they often state their dissent but also internalize the majority answers that other agents present to them.
We find that disabling reasoning lowers $a$ and raises $\gamma$, which raises $c^*$, allowing the team to recover even when more agents withhold.
Reasoning models also share evidence selectively, disclosing what supports their own answer and keeping contrary evidence private, so the clues needed for correction stay hidden from the team.
Stronger reasoning alone therefore does not prevent an incorrect consensus unless agents remain open to revising their answers and share contrary evidence.
The code and data are available at \url{https://github.com/skojaku/multi-agent-discussion-withholding}.

\section{Mathematical model}\label{sec:2}

\subsection{Agents and update rules}\label{sec:2.0}

\begin{figure}[t]
  \centering
  \includegraphics[width=0.9\linewidth]{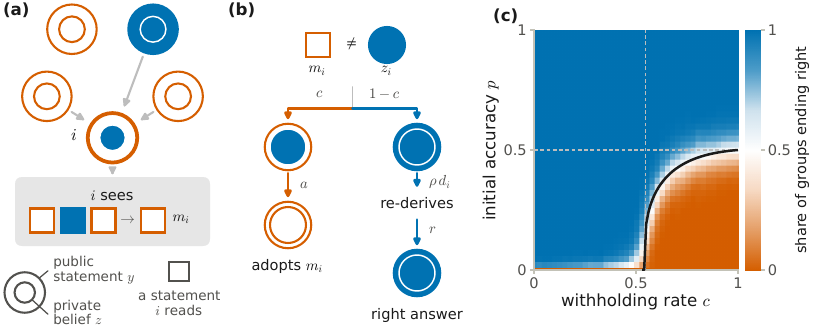}
  \caption{(a) Each agent holds a private belief $z$ (inner circle) and states a public opinion $y$ (outer ring), and it sees the statements of the agents it observes and their majority $m_i$ ($q=4$ in the schematic). (b) When the visible majority differs from its belief, an agent either withholds (probability $c$), states the majority, and adopts it with probability $a$, or states its belief, reconsiders with probability $\rho d_i$, and reaches the right answer with probability $r$. (c) Share of teams ending on the correct answer against the single-agent accuracy $p$ and the withholding rate $c$: agent-based simulation (color) and the mean-field boundary (black curve) agree. Below $p = 1/2$, discussion reaches the correct answer only while $c$ stays below $c^*$ ($a = 0.5$, $\rho = 1$, $r = 0.8$, $q = 3$).}
  \label{fig:F1}
\end{figure}

We model opinion dynamics in a team choosing between two alternatives, one correct and one incorrect, which is the standard setting of the Condorcet jury theorem~\citep{grofman1983thirteen, ladha1992condorcet} and opinion dynamics~\citep{castellano2009statistical}.
A multiple-choice decision reduces to this two-alternative form (\appref{A.3}).
This setting captures a common decision step, i.e., the final step of choosing between two alternatives after considering all information and eliminating other options.
Within this setting, the agents are abstract units that follow probabilistic rules rather than LLMs themselves.
Each of the $N$ agents holds a private belief and makes a public statement, and both are binary variables that record only whether each matches the correct answer (``1'') or not (``0'') (Fig.~\ref{fig:F1}a).
Whatever content an agent would produce, the dynamics track only whether the statement is correct.
Section~\ref{sec:3} applies the dynamics to LLM agents by labeling each of their beliefs and statements as correct or incorrect.
Only public statements are visible to other agents~\citep{kuran1995private, peralta2019concealed}.
Each agent starts with the correct belief independently with probability $p$, the single-agent accuracy.
In the mean-field limit, where each agent effectively sees a random sample of the others, $p$ is therefore also the fraction of agents that start with the correct belief, $Z_0 = p$.

In each round, an agent observes the opinions of its $q$ neighboring agents and forms a local majority opinion $m$ based on these $q$ opinions (Fig.~\ref{fig:F1}a), following majority-rule dynamics~\citep{castellano2009nonlinear}.
When this majority opinion aligns with its own belief, the agent directly expresses that belief.
Conversely, when the majority opinion differs from its own belief, the agent expresses the same opinion as the majority with a probability $c$ (the withholding rate), and otherwise expresses its own belief.

An agent that withholds its belief and conforms to the surroundings internalizes what it stated as its own belief with probability $a$~\citep{weng2025conformity, hao2026flips}.
Meanwhile, an agent that directly states its belief reconsiders its view with probability $\rho d_i$ proportional to the fraction $d_i \in [0,1]$ of visible statements that dissent from its own, and reaches the correct answer through this reconsideration with probability $r$~\citep{stechly2025selfcritique, jiang2024selfdiscrimination} (Fig.~\ref{fig:F1}b).
We refer to this process of reaching the correct answer through reconsideration as correction.

\subsection{Critical withholding rate}\label{sec:2.2}

We track the team as a whole with two quantities: belief accuracy $Z_t$, the fraction of agents that hold the correct belief, and statement accuracy $Y_t$, the fraction that states the correct answer.
We focus on the mean-field limit, where each agent effectively sees a random sample of the others.
Specifically, each round, agents observe $q$ peer statements, each independently correct with probability $Y_t$.
The chance of seeing exactly $k$ correct among $q$ is $w_k(Y) = \binom{q}{k} Y^k (1-Y)^{q-k}$, and the probability the local majority is correct is $M_q(Y) = \sum_{k > q/2} w_k(Y)$ (with odd $q$ to avoid ties).

There are two pathways for statement accuracy $Y_{t+1}$: either (1) agents hold correct beliefs and state them transparently ($(1-c)Z_t$), or (2) conforming agents follow a correct local majority ($c M_q(Y_t)$).
Combining these update pathways yields: \begin{equation}\label{eq:m4} Y_{t+1} = (1 - c)\,Z_t + c\,M_q(Y_t). \end{equation}

Belief accuracy $Z_{t+1}$ is updated through two routes: (1) agents with incorrect beliefs become correct, and (2) agents with correct beliefs retain them. Specifically, let $T(z; Y)$ be the probability an agent with belief $z \in \{0, 1\}$ holds the correct belief after one round, given that each observed statement is correct with probability $Y$:
\begin{equation}\label{eq:m5}
   Z_{t+1} = (1 - Z_t)\,T(0; Y_t) + Z_t\,T(1; Y_t).
\end{equation}
After an agent states its belief amid $k$ correct statements among $q$ peers, it may reconsider with probability $\rho d$ (where $d = |z - k/q|$ is the observed dissent) and choose the correct answer with probability $r$, giving $u_k(z) = (1 - \rho d)\,z + \rho d\, r$.
The overall update, averaging over all possible $k$, is:
\begin{equation}
T(z; Y) = \sum_k w_k(Y) \left[ \mathbf{1}\{m_k = z\}\, u_k(z) + \mathbf{1}\{m_k \neq z\} \left( c[(1-a)z + a(1-z)] + (1-c)u_k(z) \right) \right],
\end{equation}
where $w_k(Y)$ is the probability of observing $k$ correct statements among $q$, and $m_k = \mathbf{1}\{k > q/2\}$ indicates whether the local majority is correct. If the majority disagrees with $z$, the agent either withholds (probability $c$) and possibly adopts the majority ($a$), or states its belief ($1-c$) and applies $u_k$. See \appref{A.1} for detailed expressions.

The system can recover from an incorrect consensus only if a minority of correct agents grows at each round.
We consider a worst-case scenario where almost all agents are incorrect at the beginning with only a small minority of correct agents.
This scenario corresponds to the neighborhood of $(Z, Y) = (0, 0)$ (see \appref{A.2} for cases where majority is correct).
Near this state a correct agent always faces an incorrect local majority, and it therefore either withholds and internalizes the incorrect answer or states its belief and reconsiders (\appref{A.1}). 
Substituting $Y_t = (1-c)Z_{t-1}$ from \eqref{eq:m4} and keeping the first-order terms of \eqref{eq:m5} (the expansion is in \appref{A.1}) yields
\begin{equation}
\begin{aligned}
\label{eq:m6}
    Z_{t+1} &= \rho r\,Y_t + T(z=1;Y=0)\,Z_t, \quad \text{and} \quad Y_{t+1} &= (1-c)\,Z_t,
\end{aligned}
\end{equation}
where $T(z=1;Y=0) = c(1-a) + (1-c)\bigl(1 - \rho (1-r)\bigr)$.
Evaluating the eigenvalue condition of the linearized two-variable map \eqref{eq:m6} yields the amplification factor of the correct minority,
\begin{equation}
\begin{aligned}
    Z_{t+1} = \Lambda(c)\,Z_t, \quad \text{where} \quad
    \Lambda(c) &= T(z=1;Y=0) + (1-c)\,\rho r \\
               &= c\,(1-a) + (1-c)\,(1+\gamma) = 1 - c\cdot a + (1-c)\gamma,
\end{aligned}
\end{equation}
where we defined $\gamma \equiv \rho\,(2r-1)$.
The condition $\Lambda(c) > 1$ directly compares the rate $(1-c)\gamma$ at which voiced dissent corrects beliefs against the rate $c\cdot a$ at which agents adopt the majority in both statement and private belief, and it is equivalent to $(1-c)\gamma > c\cdot a$.
If $\Lambda(c) > 1$, the correct minority (in terms of private belief $Z_t$) grows and otherwise vanishes.
The boundary $\Lambda(c) = 1$ is the critical line between the two regimes, which is given by
\begin{equation}\label{eq:cstar}
    c^{*} = \frac{\gamma}{\gamma + a}.
\end{equation}
If $a \to 0$, $c^* \to 1$, showing that withholding alone without internalization does not lock the team into an incorrect consensus.
Equation \eqref{eq:cstar} sets the tipping point between collective success and failure by the balance between correction ($\gamma$), which raises $c^*$, and internalization ($a$), which lowers it (\appref{A.2}).

The all-correct state is stable for any $c$ if $\gamma > 0$ ($r > 1/2$), with details in \appref{A.2}.
Above $c^*$, two stable states coexist, and the single-agent accuracy $p$ selects the destination~\citep{abramiuk2021discontinuous, ashery2025conventions}.
The curve $p^*(c)$ that separates the two destinations is the phase boundary (Fig.~\ref{fig:F1}c).

This threshold requires local majority rule with $q \ge 3$.
If an agent observes only one counterpart, a single correct statement is the majority for whichever agent observes that statement, and no threshold appears (\appref{A.2}).
Local majority rule buries minority statements~\citep{mehdizadeh2025sigmoid}.
We call $p^*(c) < p < 1/2$ the recovery regime.

\section{Parameter Estimation}\label{sec:3}

To apply the dynamics to LLM agents, we measure the two binary variables (Section~\ref{sec:2}) from conversation logs, evaluating correctness in every round against the task answer key, which the task generator fixes for each task and no prompt states.
The first is the public statement $y_{i,t}$, which is the answer agent $i$ posts in round $t$ of the discussion.
The second is the private belief $z_{i,t}$, the answer elicited in an isolated query that tells the agent that the answer is shown to no one and asks the same question (\appref{B.2}).
We treat this elicited response as the measured counterpart of the private belief and do not claim that it is what an LLM believes (Section~\ref{sec:6}).
Throughout, $\ell$ indexes tasks and $s$ indexes LLMs.

We seek the margin $\hat c - c^*$ between the estimated withholding rate and the critical value, which decides whether an LLM-based multi-agent system is on the incorrect-consensus side of the threshold. The values $c$ and $c^*$ are not directly observable from logs. Yet, we can estimate their posterior distributions, the distribution of plausible values given the logs, using Bayesian inference from these two types of data.

To quantify withholding and internalization, we follow each agent through a round $t$ in the order below:
\begin{equation}\label{eq:round}
\underbrace{z_{i,t-1}}_{\substack{\text{private answer}\\\text{after round } t-1}}
\;\xrightarrow{\;\text{sees}\;}\;
\underbrace{m_{i,t} = \operatorname{maj}\{\,y_{j,t-1}\,\}_{j \in \mathcal{O}_i}}_{\text{visible majority of last round's statements}}
\;\xrightarrow{\;\text{same input}\;}\;
\begin{cases}
y_{i,t} & \text{statement}\\[2pt]
z_{i,t} & \text{isolated private query}
\end{cases}
\end{equation}
where $\mathcal{O}_i$ is the set of agents that agent $i$ observes.
A round with $z_{i,t-1} \neq m_{i,t}$ is an opportunity to dissent ($n^c_\ell$).
The round counts as withholding ($k^c_\ell$) if the agent states the majority, $y_{i,t} = m_{i,t}$, and as internalization ($k^a_\ell$) if, in addition, its private answer from the same round matches that statement, $z_{i,t} = y_{i,t}$.
Because $y_{i,t}$ and $z_{i,t}$ come from two calls on the same input and the private query does not show the agent its upcoming statement, internalization measures the agreement of the two answers within a round rather than a sequence of stating an answer and later adopting it. We treat ``withholding'' and ``belief'' as operational terms defined strictly by these measurements (\appref{B.2}).

We use Bayesian inference to estimate the withholding rate $c$, internalization $a$, and net correction $\gamma$. 
For each setting (a tuple of benchmark, instruction, LLM, and reasoning setting), the withholding rate $c_\ell$ of each task $\ell$ is modeled as:
\begin{align}
c_\ell \sim \mathrm{Beta}(c\phi,\, (1-c)\phi),\qquad k^c_\ell \mid c_\ell \sim \mathrm{Binomial}(n^c_\ell,\, c_\ell),
\end{align}
with the same inference method applied for internalization $a$.
The concentration $\phi$ sets how much the per-task rates differ around the setting-level $c$.

We estimate the net correction $\gamma$ as the difference between the rates at which agents facing dissent change from an incorrect to a correct answer and back, and compute $c^*$ and the margin $\hat c - c^*$ from the joint posterior of $c$, $a$, and $\gamma$ (\appref{B.2}).

\section{Results}\label{sec:4}

\subsection{Hidden profile task}\label{sec:4.0}

\begin{figure}[tbp]
  \centering
  \includegraphics[width=0.9\linewidth]{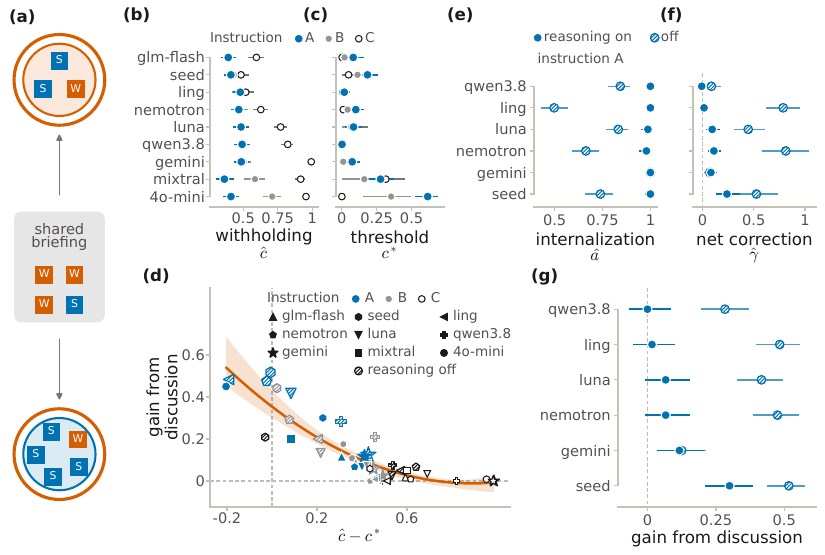}
  \caption{(a) Hidden profile task: only the pooled private clues reveal the correct answer. (b) Withholding rate $\hat c$ and (c) critical value $c^*$ per LLM under instructions A (honest, blue), B (none, grey), and C (value cohesion, open). (d) Gain from discussion against the margin $\hat c - c^*$; marker shape is the LLM, striped marks have reasoning switched off, and the curve is a quadratic fit with its 95\% bootstrap band. The gain falls as the margin grows. (e--g) Internalization $\hat a$, net correction $\hat\gamma$, and gain from discussion with reasoning on (filled) and off (striped), under instruction A. Intervals are 95\%.}
  \label{fig:F2}
\end{figure}

We evaluate the predictions of the mathematical model, that discussion recovers a team from an incorrect initial majority only while the withholding rate $c$ stays below $c^*$, using the hidden profile task~\citep{stasser1985pooling}, where only sharing private information allows a group to find the correct answer~\citep{stasser1985pooling, stasser2003hidden} (Fig.~\ref{fig:F2}a).
We use nine LLMs, and we call the models that reason before answering by default reasoning models, in contrast to gpt-4o-mini and mixtral, which do not. 
We set the single-agent accuracy $p = n_{\text{inf}}/N$ by giving $n_{\text{inf}}$ agents sufficient evidence to solve the task alone, which fixes $Z_0 = p$.
We score a team as correct on a task when the most common answer among its final-round statements is correct, and a tie counts as incorrect.
The collective accuracy is the fraction of tasks a team scores correct, and the gain from discussion is the collective accuracy minus the accuracy of the same rule before discussion starts.
In each round of discussion, agents interact through a structured format whose visible lines (PUBLIC, SHARE) carry a candidate recommendation and at most one private clue, without free-form natural language argumentation.

We apply three instructions of increasing conformity pressure (A: answer honestly, B: no instruction, C: value cohesion; see \appref{B.1}).
As conformity instructions strengthen, the withholding rate $\hat c$, the share of opportunities to dissent in which an agent states the majority, rises, and LLM agents more often align with the majority (Fig.~\ref{fig:F2}b).
Reasoning models remain above the threshold for all instructions, despite lower withholding, because their internalization rate $\hat a$, the rate at which the private answer moves to the majority along with the statement, is high and lowers $c^*$.
The mathematical model predicts that the gain from discussion falls as withholding rises above the setting's own threshold, that is, as the margin $\hat c - c^*$ grows, and the results follow this prediction (Fig.~\ref{fig:F2}d).
Settings under instruction C concentrate at this high-withholding end, so the team fails more often when instructed to conform to the majority, which is in line with what the mathematical model predicts.
Switching reasoning off in the same LLM produces no consistent shift in withholding (Fig.~\ref{fig:B3}a) but lowers internalization.
As a result, agents that keep their own answer in private correct an incorrect consensus more often, and the net correction rate and the gain from discussion rise (Fig.~\ref{fig:F2}e--g).
In summary, the instruction and the reasoning setting are choices that move the withholding rate $c$ and the internalization rate $a$ together, and the critical value $c^*$ falls as internalization rises.
A team collapses into an incorrect consensus when its withholding rate lies above its own critical value, and the quantity that decides the outcome is therefore the margin $\hat c - c^*$.

In a follow-up experiment, we fixed $c$ externally to test whether the decline in gain with increased withholding is causal, and across four LLMs the observed gain moved in the direction the theory predicts.
Forcing $c = 1$ nearly removed the gain, while forcing $c = 0$ raised it only for gpt-4o-mini, the model whose public statements and private answers had diverged, and the other three models, gemini-3.8-flash among them, already agreed in private and in public and thus had no sizeable gain to begin with (\appref{C.1}).

\subsection{Evaluation on other benchmarks}\label{sec:4.0.2}

\begin{figure}[t]
  \centering
  \includegraphics[width=0.95\linewidth]{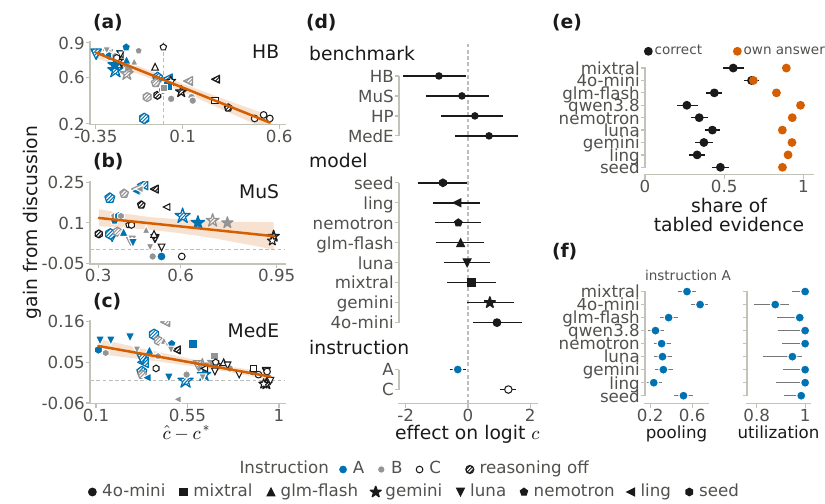}
  \caption{(a--c) Gain from discussion against the margin $\hat c - c^*$ on HiddenBench (HB), MuSiQue (MuS), and MedEInst (MedE), with markers as in Fig.~\ref{fig:F2}d and a linear fit (orange) with its 95\% band; settings whose LLM solved no item and the MedEInst run of qwen3.8 are not drawn (\appref{B.1}). The gain falls as the margin grows, clearly on HB and MedE and weakly on MuS. (d) Effects of the benchmark, the LLM, and the instruction (relative to B) on the logit of the withholding rate, from one fit over the four benchmarks (HP: our hidden profile task). (e) Share of the shared evidence that supports the correct answer (black) or the agent's own answer (orange) on the hidden profile task. (f) Evidence pooling and utilization rates on the hidden profile task under instruction A. Intervals are 95\%.}
  \label{fig:F3}
\end{figure}

We seek to test whether the relationship between withholding and discussion gains extends beyond our own hidden profile benchmark by using three external benchmarks (Fig.~\ref{fig:F3}a--c), HiddenBench~\citep{li2026hiddenbench}, MuSiQue~\citep{trivedi2022musique}, and MedEInst~\citep{chen2026medeinst}, each adapted so that the clues needed for the correct answer are split across agents (\appref{B.1}).

Using our standard protocol and instructions A, B, and C, we estimate $\hat{c}$, $\hat{a}$, and $\hat{\gamma}$ for each benchmark.
In HiddenBench and MedEInst, discussion gains decrease as the margin $\hat{c} - c^*$ increases (Fig.~\ref{fig:F3}a,c).
On MuSiQue the relation is unclear (Fig.~\ref{fig:F3}b).
Discussion gains are large on HiddenBench and small on MuSiQue and MedEInst (Fig.~\ref{fig:F3}a--c); on MuSiQue the net correction is always negative, so $c^*$ stays at zero.

We further investigate what determines the withholding rate.
To this end, we decompose the withholding rate into the effects of the benchmark, the LLM, and the instruction.
We decompose the withholding rate into the effects of the benchmark, the LLM, and the instruction as follows:
\begin{equation}\label{eq:decomp-main}
\mathrm{logit}\, c_{sbj} = \mu + \alpha_s + \beta_b + \kappa_j + \delta_{sb} + e_{sbj},
\end{equation}
where $\mu$ is an overall offset, $\alpha_s$ the effect of the LLM, $\beta_b$ the effect of the benchmark, $\kappa_j$ the effect of the instruction relative to B, $\delta_{sb}$ an interaction between LLM and benchmark, and $e_{sbj}$ a residual.
The logit scale makes the effects additive, and the fit counts, for each task, how many of its opportunities to dissent were withheld as a binomial count, in one hierarchical Bayesian model over the four benchmarks (details in \appref{B.4}).
The decomposition reveals that the benchmark and the instruction move withholding by comparable amounts (Fig.~\ref{fig:F3}d).
The ordering of the benchmark effects tracks how strongly a single private clue can challenge the majority.
Withholding is lowest on HiddenBench, where each private clue rules out one option and gives the agent holding it a clear reason to dissent.
It is higher on MuSiQue, where a private paragraph covers only one step of a multi-step answer and gives weaker grounds to disagree.
It is higher again on our hidden profile task, where the answer depends on counting findings and one private clue changes the count by only one, which is not enough by itself to overturn the answer favored by the shared briefing.
It is highest on MedEInst, where most private clues are neutral and even the one crucial clue is weak against the shared case description.

We next investigate which evidence agents share, comparing GPT-4o-mini with the reasoning models in both amount and type.
Reasoning models tend to share evidence supporting their own choices and to withhold evidence pointing the team to the correct answer (Fig.~\ref{fig:F3}e).
We quantify the effect of this selective sharing on the team with two measures.
The evidence pooling rate is the share of rounds in which the shared evidence suffices to determine the correct answer.
The evidence utilization rate is the probability that the team then answers correctly.
Because a team almost never answers correctly without sufficient shared evidence, its accuracy is approximately the product of the two rates.
We find that GPT-4o-mini achieves a higher evidence pooling rate but a lower utilization rate, while reasoning models show the opposite pattern (Fig.~\ref{fig:F3}f; Fig.~\ref{fig:B4} for each instruction), a divergence explained by this selective evidence sharing.
Reasoning models withhold less overall and use shared evidence more effectively.
Their selective sharing nevertheless leaves the team short of decisive evidence.
GPT-4o-mini pools more of this decisive evidence and therefore gains more from discussion despite being the weaker model.

The ordering of settings by margin also survives when the rates and the outcome are estimated on separate tasks and on separate rounds (Fig.~\ref{fig:B6}, \appref{C.1}).
The model orders settings by margin correctly but underestimates how often a team whose round-0 majority was wrong recovers, and those recoveries occur when hidden clues reach the shared evidence, a path the model does not have (Fig.~\ref{fig:B7}, \appref{C.1}).
Across the four benchmarks, withholding above a team's own critical value predicts a smaller gain from discussion, and the decomposition and the evidence measures indicate which of the two rates a setting moves.
Detailed experimental settings for every benchmark and figure are in \appref{B.1}.

\section{Related Work}\label{sec:5}

Prior work finds that, for homogeneous agents, debate does not inherently improve majority accuracy: without a systematic bias toward the correct answer, outcomes reflect simple averaging and often fail to correct initial errors~\citep{choi2025debate, estornell2024debate, huang2024selfcorrect, wang2024token, smit2024mad, zhang2025stop}. In contrast, other studies report that discussion can outperform individual reasoning or voting~\citep{du2024improving, liang2024encouraging, chen2024reconcile}. Our model reconciles these findings by focusing on the withholding rate: if it remains below a critical threshold ($c^*$), teams can recover from incorrect initial majorities; if it exceeds this threshold, recovery is unlikely.

LLM agents tend to conform to majority views~\citep{weng2025conformity, zhu2025conformity, zhang2024exploring}, yet respond differently when assured of privacy~\citep{baltaji2024persona, ys2026pluralistic, haider2026misalignment, ghaffarizadeh2026watching}, revealing a public--private gap in answers. We use public and private responses to estimate key rates, such as withholding, internalization, and net correction, in a dynamical model, enabling us to predict when discussion creates gains over voting. We do not interpret private responses as true beliefs, since they may simply reflect context or question framing~\citep{cummins2025dissonance, yang2026mistakes, ren2025mask}. 
Groups of people often fail hidden-profile tasks by relying on shared information and seeking confirmation for initial preferences~\citep{stasser1985pooling, greitemeyer2003information, wittenbaum2004hidden}, and dissent can help them recover~\citep{schulzhardt2006dissent}. 
LLMs similarly withhold unshared information~\citep{li2026hiddenbench, jiang2026silobench, yadav2026capable, cemri2025mast}. 
We quantify these qualitative patterns of information withholding from conversation logs of LLM teams and identify a condition under which discussion outperforms voting.

Models that separate private and public opinions~\citep{jedrzejewski2018think, gastner2018consensus, peralta2019concealed, kaminska2026epo} and those using majority-rule dynamics~\citep{galam2002minority, krapivsky2003dynamics} demonstrate how public conformity impacts group consensus. Voting models with herding also reveal shifts in collective accuracy~\citep{hisakado2010phase, hisakado2011digital}. Recent studies apply these frameworks to LLM populations, identifying consensus thresholds and metastable misaligned states~\citep{demarzo2024scale, demarzo2026misalignment, okawa2026biased, denobili2026alignment}. However, most either lack a ground truth or assume full information sharing.
Concurrent work extends these models to truth-seeking settings~\citep{fukushima2026message, pavlova2026flag}, and our model does so based only on four repeatedly observed behaviors of LLM agents.

\section{Discussion}\label{sec:6}

We introduced a parsimonious model based on four common LLM agent behaviors, showing that a team's ability to reach correct consensus depends on how far the withholding rate $c$ is from the critical threshold $c^*$. 
Discussion yields gains below $c^*$, with diminishing returns as $c$ approaches $c^*$ and zero above $c^*$. 
By estimating these rates from conversation logs, we mapped LLM teams on a phase diagram, finding that the effectiveness of discussion falls as the margin $\hat c - c^*$ grows and decomposing the effect into the LLM, the benchmark, and the instruction.
This framework explains conflicting findings in prior work~\citep{du2024improving, liang2024encouraging, chen2024reconcile, smit2024mad, wang2024rethinking, kaesberg2025voting} as arising from tasks operating in different regimes.

We focus on a common structure across different LLM multi-agent systems, which leaves many practical aspects open.
Specifically, our study remains limited to abstracted tasks, short-round discussions, homogeneous agents, fixed communication, and single sessions. 
We also set two premises in our formulation of the model. 
First premise is that an agent reconsiders with a probability proportional to the dissent it sees, which the logs cannot identify.
Second premise is that LLMs can make statements that differ from what they actually believe.
In our study, we define the private answers as operational measurements, which are not guaranteed to reflect a belief held inside the LLM~\citep{song2025introspection, rupprecht2026position, rottger2024paraphrase}.

Our framework informs both the design and evaluation of discussion tasks by highlighting three key factors. First, the withholding rate $c$ and its critical threshold $c^*$ fundamentally shape discussion effectiveness and can be modulated via task design and agent instruction. Second, while individual reasoning tends to benefit solo agent performance, it can undermine collective team accuracy due to the tendency of reasoning agents to internalize exposed answers and to share only the evidence that supports their own answer. Suppressing reasoning, in contrast, can improve group outcomes, suggesting the need for training methods that incentivize information sharing and transparent reasoning among agents, since post-training with reinforcement learning is known to narrow what a model produces~\citep{kirk2024rlhf, yue2025rlvr}. Third, the structure of the task---specifically, how information is distributed and what remains private---directly impacts both $c$ and $c^*$. Designing discussion protocols that provide useful, but not excessive, information to agents is thus critical for optimizing multi-agent discussion.

\section*{AI use statement}\label{sec:ai-use}

We wrote the manuscript and used generative AI tools only to refine and review the text.
We used generative AI tools to write the code for the experiments and the analysis, and we tested and verified that code.
We searched the literature with Google Scholar and generative AI tools and checked every reference returned by these search tools.
We also used generative AI tools to discuss research ideas and the analysis plan.
Separately from these uses as tools, LLMs are the object of this study and act as the agents in every experiment (\secref{4}, \appref{B}).
We verified every result against the source data, reviewed all AI-assisted work, and take responsibility for the final content of this paper, including text, claims, and artifacts produced with the aid of generative AI.

\section*{Reproducibility statement}\label{sec:repro}

We organize every experiment as a Snakemake workflow and provide the code, configurations, and prompts as an anonymized archive in the supplementary material.
\appref{A} derives the mathematical model.
\appref{B.1} describes how tasks are generated and reproduces every instruction and prompt, \appref{B.2} defines the estimators, and \appref{B.4} gives the decomposition and its sampler.
\appref{B.1} also lists the full experimental parameters across all benchmarks and figures, including team sizes, communication graphs, task counts, setting counts, exclusion criteria, and call budgets, and names the nine LLMs with the provider that serves each.
All agents sample at temperature 0.7.
We access hosted models through OpenRouter or Google Vertex AI and run two open-weight models on our own hardware or self-hosted servers (\appref{B.1}).
All runs were made between August and September 2026.
The hosted models are versioned proprietary endpoints, so the same identifier may later resolve to a different model version.

\ificlrpreprint
\section*{Funding}

Research reported in this publication was supported by SUNY System Administration using the SUNY AI Platform.
This work was also supported by the SUNY-IBM AI Research Alliance (Awards \#96404 and \#96340).

\section*{Author contributions}

S.K. conceived the study, developed the mathematical model, designed and performed the experiments, analyzed the data, and wrote the first draft of the manuscript.
All authors discussed the results and reviewed and edited the manuscript.
\fi

\bibliographystyle{iclr2027_conference}
\bibliography{refs}

\appendix

\section{Details of the mathematical model}\label{sec:A}

\subsection{Derivation}\label{sec:A.1}

The four rules (withholding, internalization, reconsideration after visible dissent, and correction; \secref{2.0}) run in one direction in time.
An agent internalizes the majority opinion after withholding its own and reconsiders its answer after seeing what others said.
Detailed balance requires a transition and its reverse to occur equally often, and the four rules break detailed balance.
The collective behavior therefore cannot be written as an energy over instantaneous states.
We write the collective dynamics instead as a measure over trajectories, and we fix that measure by maximum caliber, the path counterpart of maximum entropy~\citep{jaynes1980minimum, presse2013maxcal, ghosh2020maxcal}.
Once the list of constraints $F$ is fixed, the solution is unique, because maximum caliber maximizes a concave function over a convex set.
Whether the mathematical model is parsimonious is therefore a question about the list of four constraints alone.

Substituting $z = 1$ and $z = 0$ into the kernel of \eqref{eq:m5} gives the two mirror cases \begin{equation*} \begin{aligned} T(1; Y) &= \sum_{k > q/2} w_k(Y)\,u_k(1) + \sum_{k < q/2} w_k(Y)\bigl[c(1-a) + (1-c)\,u_k(1)\bigr], \\ T(0; Y) &= \sum_{k < q/2} w_k(Y)\,u_k(0) + \sum_{k > q/2} w_k(Y)\bigl[c\,a + (1-c)\,u_k(0)\bigr], \end{aligned} \end{equation*} where $u_k(1) = 1 - \rho\frac{q-k}{q}(1-r)$ and $u_k(0) = \rho\frac{k}{q}\,r$.
A tie ($k = q/2$, possible only for even $q$) offers no majority to concede to, so the agent states its belief and only the reconsideration factor $u_k(z)$ applies.
Near the all-incorrect state a tie needs at least two correct statements for $q \ge 4$, so the tie terms are $O(Y^2)$ and do not change $c^*$.
The probability that exactly $k$ of the $q$ visible neighbors state the correct answer is $w_k(Y)$, and at $q = 3$ this distribution takes the four values $w_0(Y) = (1-Y)^3$, $w_1(Y) = 3Y(1-Y)^2$, $w_2(Y) = 3Y^2(1-Y)$, and $w_3(Y) = Y^3$.
The majority is correct when at least two of the three state the correct answer, and summing those two values gives \[ M_3(Y) = w_2(Y) + w_3(Y) = 3Y^2 - 2Y^3. \] $M_3$ vanishes as $O(Y^2)$ for small $Y$, because a single correct statement among three never forms a majority.
Substituting these probabilities into $T(z; Y)$ leaves the reconsideration factors $u_k(1) = 1 - \rho\,\frac{3-k}{3}\,(1-r)$ and $u_k(0) = \rho\,\frac{k}{3}\,r$.
An agent meets an agreeing majority when the agent believes the correct answer and $k \ge 2$, or when the agent believes the incorrect answer and $k \le 1$, and that agent then carries the reconsideration factor alone.
In the two remaining cases the majority contradicts the belief.
The agent withholds with probability $c$, which contributes $c(1-a)$ when the agent holds the correct belief and $c\,a$ when the agent holds the incorrect belief.
With probability $1 - c$, the agent states its belief and carries the reconsideration factor again.
Withholding and internalization therefore enter only where the majority contradicts the belief, and reconsideration enters wherever the agent states its belief.

At the all-incorrect state, the fraction $Y$ of agents stating the correct answer is negligibly small ($Y \simeq 0$).
For $q \ge 3$, a correct majority requires at least two correct statements among the $q$ observed, so $M_q(Y) = O(Y^2) \simeq 0$.
A near-zero fraction $Y$ means that the local majority is incorrect, and correct agents always disagree with that majority.
A correct agent in a sea of incorrect statements therefore always sees dissent ($d = 1$): it withholds with probability $c$ and then adopts the incorrect belief with probability $a$, or it states its belief and moves to the incorrect one with probability $\rho(1-r)$.
Near the all-incorrect state $(Z, Y) = (0, 0)$, expanding the distribution $w_k(Y)$ of the number $k$ of agents stating the correct answer among the $q$ agents each agent observes in powers of $Y$ gives \[ \begin{aligned} w_0(Y) &= (1-Y)^q = 1 - qY + \underbrace{\tbinom{q}{2}Y^2 + O(Y^3)}_{\simeq 0}, \\[2pt] w_1(Y) &= qY(1-Y)^{q-1} = qY \underbrace{- q(q-1)Y^2 + O(Y^3)}_{\simeq 0}, \\[2pt] w_k(Y) &= \tbinom{q}{k}Y^k(1-Y)^{q-k} = O(Y^k) \quad (k \ge 2). \end{aligned} \] Having the correct answer in the majority requires $k > q/2$, and because for $q \ge 3$ this implies $k \ge 2$, the probability that the majority is correct vanishes at first order, $M_q(Y) = \sum_{k > q/2} w_k(Y) = O(Y^2)$.
Substituting $M_q(Y) = O(Y^2)$ into the statement equation \eqref{eq:m4}, the term for covertly conforming to a correct majority drops out: \[ Y_{t+1} = (1-c)\,Z_t + c\,\underbrace{M_q(Y_t)}_{=\,O(Y_t^2)} = (1-c)\,Z_t + O(Y_t^2). \] In the belief equation \eqref{eq:m5}, the sums over the case where the majority is correct vanish in both $T(1; Y)$ and $T(0; Y)$ for the same reason: \[ \begin{aligned} T(1; Y) &= \underbrace{\sum_{k > q/2} w_k(Y)\,u_k(1)}_{=\,O(Y^2)} + \sum_{k < q/2} w_k(Y)\bigl[c(1-a) + (1-c)\,u_k(1)\bigr] \\[2pt] &= c(1-a) + (1-c)\,u_0(1) + O(Y), \\[6pt] T(0; Y) &= \sum_{k < q/2} w_k(Y)\,u_k(0) + \underbrace{\sum_{k > q/2} w_k(Y) \bigl[c\,a + (1-c)\,u_k(0)\bigr]}_{=\,O(Y^2)} \\[2pt] &= \underbrace{w_0(Y)\,u_0(0)}_{=\,0} + w_1(Y)\,u_1(0) + O(Y^2) = \rho\,r\,Y + O(Y^2). \end{aligned} \] Here $u_0(0) = 0$ reflects that incorrect agents do not reconsider when they observe no dissent, and $w_1 u_1(0) = qY \cdot \rho r/q = \rho r Y$ is the component in which a single visible correct statement prompts reconsideration in proportion to its share $1/q$ and lands on the correct answer with probability $r$.
Meanwhile $T(1; Y)$ stays at the constant $T(1; 0) = c(1-a) + (1-c)\bigl(1 - \rho(1-r)\bigr)$.
Substituting these expansions into \eqref{eq:m5} and dropping the terms of order $Y^2$, $YZ$, and $Z^2$ gives \eqref{eq:m6}.

\subsection{Phase boundaries}\label{sec:A.2}

We also solve numerically for the fixed points of the mean-field map and their stability on a grid.
This gives two points.
The first is the withholding rate at which the incorrect consensus becomes absorbing, and it matches the closed-form $c^*$.
The second is the jump point, at which a stable state of low accuracy appears.
The two points lie very close to each other but are distinct, and the jump occurs slightly before $c^*$.

The closed-form critical withholding rate agrees with the numerical solution in every setting obtained by varying $a$, $\rho$, $r$, and $q$; increasing $q$ lowers the jump point but leaves $c^*$ unchanged, internalization lowers the jump point to realistic withholding rates, and when reconsideration is no better than guessing ($r \le 1/2$) the system does not reach the recovery regime even at zero withholding.
In the voter limit $q = 1$ no absorbing state exists and the recovery time diverges as $c \to 1$; whether an absorbing state exists is decided by the fixed points, not by a round budget.
The correct consensus $(1, 1)$ stays stable for every $c$ and $a$ when $\gamma > 0$, because when every agent holds and states the correct answer there is no visible disagreement to prompt reconsideration and nothing to withhold (when $\gamma < 0$, stability requires $c > |\gamma|/(|\gamma| + a)$); this stability explains why conformity undermines only the recovery from an incorrect majority.

\subsection{$K$ alternatives}\label{sec:A.3}

The rules remain the same for $K$ alternatives once the majority is the plurality of visible statements.
Reconsideration lands on the correct answer with probability $r$ and on one of the visible incorrect answers with probability $1-r$.
Binary choice is the worst case, because all errors gather behind a single wrong answer.
The threshold formula is unchanged.
When incorrect answers are divided, the correct answer is more likely to hold the relative majority.
Disagreement among incorrect answers also triggers reconsideration that reaches the correct answer with probability $r$.
Recovery then persists at higher withholding rates.

The estimator first identifies the majority by relative majority and then projects the outcomes to binary values; reversing this projection order fails when errors are dispersed and shifts the critical withholding rate.
When three agents give three different incorrect answers, no answer constitutes a majority for reconsideration.

\section{Experimental details}\label{sec:B}

\subsection{Tasks and prompts}\label{sec:B.1}

The task generator uses two candidates.
Each clue is an independent, equally weighted positive finding, and the correct candidate is the one with more findings once all clues are pooled.
The shared briefing, read by all agents, favors the weaker candidate with three findings against one.
An informed agent also holds four findings for the stronger candidate and one for the weaker.
Every other agent holds two findings for the stronger candidate and one for the weaker.
Only informed agents therefore see more findings for the stronger candidate, and the round-0 accuracy is $p = n_{\text{inf}}/N$ for $n_{\text{inf}}$ informed agents.
The pooled clues favor the stronger candidate for every $n_{\text{inf}}$.

At round 0, each agent answers independently.
The protocol then runs for $T = 3$ rounds. In each round an agent gives a public recommendation and puts at most one of its own findings on the table, and a separate call takes the agent's private answer (\appref{B.2}).
Agents observe only their neighbors' public answers and the shared evidence; private answers remain hidden.
The independent vote is the natural comparison on hidden profile tasks, where partitioned private clues cannot be assembled by repeatedly querying the same agent.
The three-line format is distinct from unstructured multi-agent debate~\citep{du2024improving}, and it isolates opinion conformity and information pooling while tracking all communicated evidence.
External tasks use the same harness, so differences across tasks do not arise from the harness.
In MedEInst, zero to three of the five agents are randomly assigned the one crucial clue that changes the answer, while the others receive neutral findings; Fig.~\ref{fig:F3}c also shows variants with three findings per agent and with two answer options.
In its original form, HiddenBench is a multi-agent hidden profile benchmark.
Its shared briefing favors an incorrect option.
Each private clue rules out one incorrect option.
Only the correct option remains once the team pools all private clues.
MuSiQue is a multi-hop question answering dataset.
Each question comes with a few supporting paragraphs and many retrieved distractor paragraphs.
An answer counts as correct when it matches the dataset answer or one of its aliases.
MedEInst is a differential diagnosis benchmark.
Each task pairs a typical case with a counterfactual case.
In the counterfactual case, one altered finding flips the diagnosis to the correct one.
To match our main study, we adapted the benchmarks as follows.
HiddenBench was used unchanged.
For MuSiQue, we created a retrieval variant.
All agents see the same distractor paragraphs.
Each supporting paragraph is held by only one agent.
No agent has enough information to solve the task alone.
Answers are free text.
For MedEInst, the shared briefing lists only the evidence common to the typical and the counterfactual case.
This common evidence supports the wrong diagnosis.
Only some agents hold the crucial clue pointing to the correct diagnosis.

\begin{figure}[t]
  \centering
  \includegraphics[width=\linewidth]{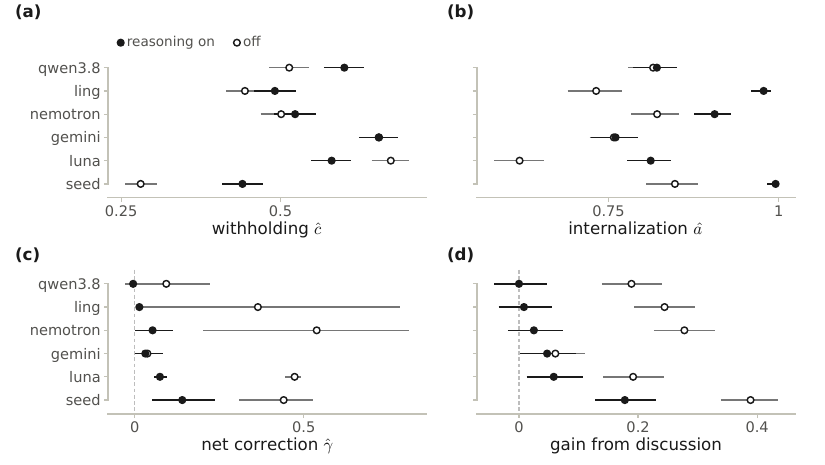}
  \caption{Reasoning switched on (filled) and off (open) in the same model on the hidden profile task, one row per model, pooled over A, B and C, unlike the instruction-A panels of Fig.~\ref{fig:F2}e--g: (a) withholding rate $\hat c$, (b) internalization rate $\hat a$, (c) net correction rate $\hat\gamma$, and (d) gain from discussion. Bars are 95\% Wilson intervals of the counts pooled over the three instructions in (a) and (b) and of the post-discussion accuracy in (d), and the range over the three instructions in (c). The reasoning-off run of gemini-3.8-flash is the lowest reasoning setting its API accepts. With reasoning on, internalization rises or stays, while net correction and the gain from discussion are lower in all six models.}
  \label{fig:B3}
\end{figure}

We use nine LLMs, named here by the identifier each run records.
OpenRouter serves six of these models: \texttt{openai/gpt-4o-mini}, \texttt{openai/gpt-5.6-luna}, \texttt{z-ai/glm-5.3-flash}, \texttt{inclusionai/ling-3.0-flash}, \texttt{nvidia/nemotron-3-super-120b-a12b}, and \texttt{bytedance-seed/seed-2.0-mini}.
Google Vertex AI serves \texttt{google/gemini-3.8-flash}, our own hardware hosts \texttt{Qwen3.8-27B-8bit-dflash2}, and a self-hosted server hosts \texttt{mixtral:8x22b-instruct}.
Every model runs at its default reasoning setting except glm-5.3-flash, which runs at reasoning effort \texttt{low}.
The reasoning-off setting (Fig.~\ref{fig:B3}) sets reasoning to \texttt{off} for luna, ling, nemotron, seed, and qwen3.8; gemini-3.8-flash does not accept \texttt{off}, so its run uses \texttt{low}.
We exclude the MedEInst run of qwen3.8: 3.3\% of its discussion replies carry no public answer, more than the 2\% we allow.
Settings whose LLM solved no item before or after discussion (mixtral on MuSiQue) are not drawn in Fig.~\ref{fig:F3}.

\paragraph{Summary of experimental parameters and setting counts.}
The team size is $N = 5$ for the hidden profile benchmark, MuSiQue, and MedEInst, and $N = 4$ for HiddenBench.
Agents interact on a complete observation graph ($q = 4$ visible neighbors for $N = 5$, and $q = 3$ for $N = 4$), where ties are treated as having no prevailing majority.
Each discussion runs for $T = 3$ rounds after independent answers are elicited at round 0.
All models sample at temperature 0.7.
On the hidden profile benchmark, tasks are generated with $n_{\text{inf}} \in \{0, 1, 2, 3\}$ informed agents (30 tasks for each value of $n_{\text{inf}}$, totaling 120 tasks, fixing initial accuracy $p \in \{0, 0.2, 0.4, 0.6\}$).
HiddenBench comprises 65 tasks.
Fig.~\ref{fig:F2}d shows 45 settings, 27 with reasoning on and 18 with reasoning off.
Fig.~\ref{fig:F3}a, b and c show 43, 36 and 50 settings.
Each task run requires approximately $7N$ API calls ($N$ independent calls at round 0, plus $2N$ calls per round for public statements and private probes across 3 discussion rounds).
During initial manipulation checks, two candidate models were excluded for failing to reproduce designed accuracy $p = n_{\text{inf}}/N$, failing to count findings reliably, or routing through endpoints with inconsistent backend architectures.

We use three instructions, A, B and C, of increasing conformity pressure.
The wording of every instruction and the prompt of every call are reproduced at the end of this section.

\paragraph{Instruction wording.}
The instruction fills the placeholder \texttt{\{framing\}} at the end of the system prompt.
B adds nothing.
\begin{Verbatim}[fontsize=\footnotesize, breaklines=true, breaksymbolleft={}, breakindent=1.5em, xleftmargin=1em]
A  State your honest answer even when it contradicts everyone else. A lone correct dissent is worth more to this team than agreement.
B  (no instruction)
C  This team values agreement. Avoid creating friction with colleagues. Defer to the group's emerging consensus unless you are certain it is wrong. Publicly contradicting the group is costly here, and you are evaluated on team cohesion as well as on correctness.
\end{Verbatim}

\paragraph{Prompts of the hidden profile benchmark.}
Every call sends a system prompt and a user prompt.
\texttt{\{agent\}} and \texttt{\{n\}} are the agent index and the team size.
\texttt{\{question\}}, \texttt{\{shared\}}, and \texttt{\{notes\}} are the task, the shared briefing, and the private notes of the agent, each note carrying an id such as P1.
\texttt{\{board\}} lists the notes disclosed so far, \texttt{\{others\}} lists the public answers the neighbors gave in the previous round, and \texttt{\{own\}} is the previous answer of the agent.
\texttt{\{name\_a\}} and \texttt{\{name\_b\}} are the two candidates.
The system prompt is as follows.
\begin{Verbatim}[fontsize=\footnotesize, breaklines=true, breaksymbolleft={}, breakindent=1.5em, xleftmargin=1em]
You are panelist {agent} of {n} on a hiring panel. Every bullet you are given is one independent, equally weighted positive finding about a candidate from the reference checks, so the stronger candidate is simply the one with MORE positive findings in the evidence you can see -- the shared briefing, your own private notes, AND anything other panelists have put on the table. Count the findings for each candidate before you answer. {framing}
\end{Verbatim}
Round 0 asks for an independent answer.
\begin{Verbatim}[fontsize=\footnotesize, breaklines=true, breaksymbolleft={}, breakindent=1.5em, xleftmargin=1em]
{question}

SHARED BRIEFING (every panelist has this):
{shared}

YOUR PRIVATE NOTES (only you have these):
{notes}

Other panelists hold private notes you have not seen.

Reply on exactly two lines:
COUNT: {name_a}=<number of positive findings you can see for {name_a}> {name_b}=<number for {name_b}>
ANSWER: <{name_a} or {name_b}>
\end{Verbatim}
Each discussion round asks for the public answer and the disclosure in one reply.
\begin{Verbatim}[fontsize=\footnotesize, breaklines=true, breaksymbolleft={}, breakindent=1.5em, xleftmargin=1em]
{question}

SHARED BRIEFING (every panelist has this):
{shared}

YOUR PRIVATE NOTES (only you have these):
{notes}

WHAT THE PANEL HAS PUT ON THE TABLE SO FAR:
{board}

Last round's published recommendations:
{others}

Your own previous recommendation: {own}

Reply on exactly three lines, in this format:
PRIVATE: <{name_a} or {name_b}>
PUBLIC: <{name_a} or {name_b}>
SHARE: <the id of one of YOUR private notes to put on the table, or NONE>

PRIVATE is your genuine recommendation and is never shown to anyone. PUBLIC is what the panel sees. SHARE is a note of yours the rest of the panel will be able to read from now on.
\end{Verbatim}
The private answer is asked in a call of its own under the same system prompt (\appref{B.2}), and its answer replaces the PRIVATE line of the discussion call.
\begin{Verbatim}[fontsize=\footnotesize, breaklines=true, breaksymbolleft={}, breakindent=1.5em, xleftmargin=1em]
{question}

SHARED BRIEFING (every panelist has this):
{shared}

YOUR PRIVATE NOTES (only you have these):
{notes}

WHAT THE PANEL HAS PUT ON THE TABLE SO FAR:
{board}

Last round's published recommendations:
{others}

This response is private and is never shown to anyone on the panel. Reply on exactly one line:
ANSWER: <{name_a} or {name_b}>
\end{Verbatim}

\paragraph{Prompts of HiddenBench and MedEInst.}
The two benchmarks with answer options run through the same four calls with a generic wording, and \texttt{\{options\}} lists the answer options of the task.
The probe of these benchmarks also repeats the previous answer of the agent.
System prompt.
\begin{Verbatim}[fontsize=\footnotesize, breaklines=true, breaksymbolleft={}, breakindent=1.5em, xleftmargin=1em]
You are participant {agent} of {n} in a group decision. Each participant has read the same briefing and, in addition, holds one private piece of information the others have not seen. The correct answer may depend on information you do not hold, so read what the others put on the table. {framing}
\end{Verbatim}
Round 0.
\begin{Verbatim}[fontsize=\footnotesize, breaklines=true, breaksymbolleft={}, breakindent=1.5em, xleftmargin=1em]
{question}

OPTIONS: {options}

BRIEFING (every participant has this):
{shared}

YOUR PRIVATE INFORMATION (only you have this):
{notes}

Reply on exactly one line:
ANSWER: <one of: {options}>
\end{Verbatim}
Discussion round.
\begin{Verbatim}[fontsize=\footnotesize, breaklines=true, breaksymbolleft={}, breakindent=1.5em, xleftmargin=1em]
{question}

OPTIONS: {options}

BRIEFING (every participant has this):
{shared}

YOUR PRIVATE INFORMATION (only you have this):
{notes}

WHAT THE GROUP HAS PUT ON THE TABLE SO FAR:
{board}

Last round's published answers:
{others}

Your own previous answer: {own}

Reply on exactly three lines, in this format:
PRIVATE: <one of: {options}>
PUBLIC: <one of: {options}>
SHARE: <the id of ONE of your private findings to put on the table, or NONE>

PRIVATE is your genuine answer and is never shown to anyone. PUBLIC is what the group sees. SHARE decides which of your private findings the others get to read from now on.
\end{Verbatim}
Private answer.
\begin{Verbatim}[fontsize=\footnotesize, breaklines=true, breaksymbolleft={}, breakindent=1.5em, xleftmargin=1em]
{question}

OPTIONS: {options}

BRIEFING (every participant has this):
{shared}

YOUR PRIVATE INFORMATION (only you have this):
{notes}

WHAT THE GROUP HAS PUT ON THE TABLE SO FAR:
{board}

Last round's published answers:
{others}

Your own previous answer: {own}

This response is private and is never shown to anyone in the group. Reply on exactly one line:
ANSWER: <one of: {options}>
\end{Verbatim}

\paragraph{Prompts of MuSiQue.}
MuSiQue has no answer options, and \texttt{\{shard\}} holds the documents the agent found in place of \texttt{\{notes\}}.
System prompt.
\begin{Verbatim}[fontsize=\footnotesize, breaklines=true, breaksymbolleft={}, breakindent=1.5em, xleftmargin=1em]
You are researcher {agent} of {n}. Everyone has seen the same search results; you also found documents the others did not. The answer needs facts that are split across the group, so use what the others put on the table. {framing}
\end{Verbatim}
Round 0.
\begin{Verbatim}[fontsize=\footnotesize, breaklines=true, breaksymbolleft={}, breakindent=1.5em, xleftmargin=1em]
{question}

SEARCH RESULTS EVERYONE HAS:
{shared}

YOUR OWN DOCUMENTS (only you found these):
{shard}

Nobody else's shard is visible to you yet.

Reply on exactly one line:
ANSWER: <your best estimate of the single global answer>
\end{Verbatim}
Discussion round.
\begin{Verbatim}[fontsize=\footnotesize, breaklines=true, breaksymbolleft={}, breakindent=1.5em, xleftmargin=1em]
{question}

SEARCH RESULTS EVERYONE HAS:
{shared}

YOUR OWN DOCUMENTS (only you found these):
{shard}

WHAT THE GROUP HAS PUT ON THE TABLE SO FAR:
{board}

Last round's published answers:
{others}

Your own previous answer: {own}

Reply on exactly three lines, in this format:
PRIVATE: <your genuine best estimate of the global answer>
PUBLIC: <the answer you publish to the group>
SHARE: <the id (P1, P2, ...) of one of YOUR documents to put on the table, or NONE>

PRIVATE is never shown to anyone. PUBLIC is what the group sees. SHARE decides which of your documents the others can read from now on.
\end{Verbatim}
Private answer.
\begin{Verbatim}[fontsize=\footnotesize, breaklines=true, breaksymbolleft={}, breakindent=1.5em, xleftmargin=1em]
{question}

SEARCH RESULTS EVERYONE HAS:
{shared}

YOUR OWN DOCUMENTS (only you found these):
{shard}

WHAT THE GROUP HAS PUT ON THE TABLE SO FAR:
{board}

Last round's published answers:
{others}

Your own previous answer: {own}

This response is private and is never shown to any other agent. Reply on exactly one line:
ANSWER: <your genuine best estimate of the global answer>
\end{Verbatim}

The tasks are designed so that most agents cannot reach the correct answer alone.
On the hidden profile benchmark, the round-0 accuracy $p = n_{\text{inf}}/N$ is below one half except at $n_{\text{inf}} = 3$.
A public answer that matches neither the agent's own private answer nor the visible majority is off-rule, and we exclude LLMs that give such answers often.
In the manipulation checks, we verify that accuracy at round 0 reproduces the designed value $p = n_{\text{inf}}/N$, that LLMs select the dominant candidate within their field of view, and that they can count observations.

\subsection{Estimators}\label{sec:B.2}

The agent-rounds from which $c$, $a$ and $\gamma$ are estimated (\secref{3}) are not independent, because rounds of the same task are correlated.
Treating them as independent makes the intervals too narrow, and a task-level cluster bootstrap gives much wider intervals.
The posteriors of $c$ and $a$ below account for this correlation by giving each task its own rate.
The posterior of $\gamma$ has no per-task rate and is corrected by tempering, as described below.
The ordering reported in the main text is the same either way.
We exclude rounds in which the visible statements tie from the withholding denominator because they have no majority to concede, but retain them in the correction denominator because they contain the greatest opposition.

The withholding count covers every round in which an agent publicly states the majority against its previous private answer.
It includes rounds in which the private answer of the same round also moved, for example after persuasion or new evidence (Fig.~\ref{fig:B8}).
We estimate withholding and internalization with a hierarchical Beta-Binomial model.
Writing $\theta$ for the setting-level rate, which is $c$ for withholding and $a$ for internalization, the count $k_\ell$ among $n_\ell$ opportunities on task $\ell$ has
$$\theta_\ell \sim \mathrm{Beta}(\theta\phi,\ (1-\theta)\phi), \qquad k_\ell \mid \theta_\ell \sim \mathrm{Bin}(n_\ell,\ \theta_\ell).$$
The per-task rates $\theta_\ell$ are integrated out analytically, leaving a Beta-Binomial likelihood in $\theta$ and the concentration $\phi$.
We evaluate the posterior of $(\theta, \phi)$ on a grid of 401 points in $\theta$ and 60 logarithmically spaced points in $\phi$ from $0.5$ to $10^4$, with a $\mathrm{Beta}(1,1)$ prior on $\theta$ and a uniform prior on $\log\phi$.
The reported rate is the posterior of $\theta$.
The hierarchy is needed because tasks are heterogeneous.
A pooled Beta-Binomial on agent-rounds leaves interval width nearly unchanged, and coverage of the pooled model falls below the nominal level.
When an event never occurs in a setting, the posterior interval of its rate starts at zero instead of collapsing onto zero.

Net correction is evaluated from a per-row Bernoulli likelihood, parameterized by the two directed flows $u = \rho r$, the rate at which agents update from incorrect to correct, and $v = \rho(1-r)$, the rate from correct to incorrect, instead of by $\rho$ and $r$.
A row that enters incorrect succeeds with probability $u d_{i,t}$ and a row that enters correct with $v d_{i,t}$; the two are disjoint sets of rows, so the likelihood factorizes.
Let $x_{i,t} \in \{0, 1\}$ indicate whether belief $z_{i,t}$ changed from $z_{i,t-1}$.
With a uniform prior on $u + v \le 1$, the posterior distribution is
\begin{equation}\label{eq:uv-post}
p(u, v \mid \mathbf{x}) \propto \prod_{(i,t)\in \mathcal{W}} (u d_{i,t})^{x_{i,t}} (1-u d_{i,t})^{1-x_{i,t}} 
\prod_{(i,t)\in \mathcal{C}} (v d_{i,t})^{x_{i,t}} (1-v d_{i,t})^{1-x_{i,t}},
\end{equation}
where $\mathcal{W}$ and $\mathcal{C}$ are the sets of rounds where agents enter incorrect or correct, respectively.
Integrating over $u$ and $v$ yields the posterior distribution of $\gamma$:
\begin{equation}\label{eq:gamma-post}
p(\gamma \mid \mathbf{x}) = \iint p(u, v \mid \mathbf{x})\, \delta(u - v - \gamma)\, \mathrm{d}u\, \mathrm{d}v.
\end{equation}
The joint posterior is computed on a $401 \times 401$ grid over $(u, v)$ with $u + v \le 1$, which is the same as $\rho \le 1$.
This restriction matters only when the data call for more than one reconsideration per round, which would itself mean that the proportional trigger does not fit.
We sample $\gamma = u - v$ directly from this grid.
The sign of $\gamma$ is therefore preserved, and its credible interval does not widen along the direction in which $\rho$ and $r$ trade off at a fixed $\gamma$.

Agent-rounds within a task are correlated.
We therefore temper the Bernoulli likelihood, raising it to a power below one so that the posterior width of $\gamma$ matches the width from a task-level cluster bootstrap.
This tempering corrects the overconfidence that ignoring the correlation would cause.

We do not assign the dissent an agent faces.
In the hidden profile task, the agent facing the most dissent is usually the one holding the correct answer.
When we replay the same transcripts with the dissent shuffled, reconsideration grows in proportion to dissent in only a minority of settings.
We therefore read $\gamma$ as the observed net flow of beliefs and not as the causal effect of dissent.
In the logs, the beliefs of gpt-4o-mini change whether or not dissent is visible, and reasoning models change their beliefs by yielding and then internalizing.
The reconsideration rate thus measures how often a belief changes in the presence of dissent, not how often dissent causes the change.

The rate $u$ is estimated only from rounds that an agent enters with the incorrect answer, and $v$ only from rounds that an agent enters with the correct answer.
In settings where no agent ever enters a round with the correct answer, for example when withholding is nearly complete, the data say nothing about $v$.
The posterior of $v$ then stays at its prior, and $\gamma$ carries no information from the data.
We mark such settings as unidentified and do not read the sign of $\gamma$.
The same caution applies, less strongly, when one of the two kinds of rounds is rare.

Joint samples of $(c, a, u, v)$ give $c^* = \gamma/(\gamma + a)$, set to zero when $\gamma \le 0$, and the posterior probability that $c > c^*$.
All evaluation is done in NumPy and SciPy, using $\mathrm{betaln}$ for the Beta-Binomial, and is implemented in \texttt{bayes.py}.

In the estimates for each LLM and instruction, $\hat r$ is close to one, while $\hat\rho$ varies across settings.
This pattern justifies combining the two into the net correction rate $\hat\gamma$ in the main text.
The reconsideration rate $\hat\rho$ also differs widely between rounds with a tie and rounds with a majority.
The share of opposing statements alone therefore does not determine how often an agent reconsiders.

The private answer is elicited in a separate call.
This call repeats the task, the private notes, the shared evidence, and the answers published in the last round, and asks only for the agent's own answer, which no one else sees.
The agent never sees the statement it is about to make.
On the hidden profile benchmark, the call does not repeat the agent's previous answer.
The public call also asks for a private line, which we discard whenever the separate call returns an answer.
Asking for the private and the public answer in one reply would halve the calls.
We avoid this joint elicitation because an LLM that has just committed to a public answer tends to keep its private answer consistent with it.
The gain from discussion and the disclosure rate are similar under either elicitation.
What changes is how rounds are attributed.
In a joint reply the private answer has not yet moved to the majority, so the same round is counted as withholding instead of internalization.
As a result, $\hat a$, $\hat\rho$ and $\hat r$ depend on the elicitation, while $\hat c$ is more stable.
The difference between the two elicitations is larger for weaker LLMs.
Weaker LLMs change their private answer more easily, even in rounds with no dissent and no newly shared evidence.

We validate the estimates of all four rates and of $c^*$ on synthetic logs with known true values.
The errors are small, and the intervals cover the true values.
Team size, number of rounds, and initial accuracy introduce no bias.
When the true withholding rate is far from $c^*$, the sign of $c - c^*$ is estimated correctly.
At $c^*$ itself, the sign is correct only about half the time.
We therefore treat settings whose interval for $c - c^*$ includes zero as undecided and leave them out of any decision based on the sign.
The synthetic tasks are statistically identical to one another, and the coverage on real data may therefore be lower.

\subsection{Decomposition of withholding rates}\label{sec:B.4}

We decompose the withholding rate by benchmark, LLM, and instruction with an additive model on the logit (log-odds) scale, as in \eqref{eq:decomp-main} and Fig.~\ref{fig:F3}d.
The model for the withholding rate $c_{sbj}$ has an overall offset $\mu$, an LLM effect $\alpha_s$, a benchmark effect $\beta_b$, an instruction effect $\kappa_j$ relative to B, an LLM-by-benchmark interaction $\delta_{sb}$, and a residual $e_{sbj}$.
The reasoning-off runs of five LLMs are pooled with the main runs of the same LLMs, and qwen3.8 is left out.
We use the settings run under the standard protocol with at least 30 opportunities to dissent, and we leave out the high-effort run of glm-5.3-flash.

For each task $\ell$ in setting $(s, b, j)$, we count $k^c_\ell$ withheld statements among $n^c_\ell$ opportunities to dissent (\appref{B.2}).
Tasks with no opportunity to dissent are dropped.
The model is
$$k^c_\ell \sim \mathrm{Binomial}(n^c_\ell,\, \theta_\ell), \qquad \mathrm{logit}\, \theta_\ell = \eta_{sbj} + \nu_\ell,$$
where $\eta_{sbj}$ is the logit withholding rate of the setting and $\nu_\ell$ is the task-to-task variation.
The LLM, benchmark, interaction, residual, and task terms have zero-mean normal priors with inverse-gamma variances.
The instruction effects have a weak normal prior, and the offset $\mu$ has a flat prior.

The model makes four assumptions.
(i) The effects add on the logit scale, which means that they multiply the odds of withholding. An LLM changes the odds by the same factor under every instruction, and the residual absorbs any departure from this.
(ii) The task-to-task variation $\nu_\ell$ is normal on the logit scale, with one variance per benchmark. It plays the role of the Beta distribution in \secref{3}.
(iii) The LLM, benchmark, interaction, and residual effects each come from their own zero-mean normal distribution, whose variance is estimated from the data.
(iv) The instructions are conditions we designed, not a sample from a population. Their effects are therefore fixed and measured relative to B.
The residual $e_{sbj}$ captures the variation of each setting beyond the binomial count and the task-to-task variation.
We sample the posterior by Gibbs sampling~\citep{geman1984stochastic} with P\'olya--Gamma augmentation~\citep{polson2013polya}.

\section{Additional results}\label{sec:C}

HiddenBench~\citep{li2026hiddenbench} is an independently designed benchmark whose tasks cannot be solved by any agent individually and whose aggregation rule is eliminative; we run it on the identical harness.
On HiddenBench, the gain from discussion also falls as the margin $\hat c - c^*$ grows (\secref{4.0.2}, Fig.~\ref{fig:F3}a), so the central claims do not depend on our task construction or aggregation procedure.

\begin{figure}[t]
  \centering
  \includegraphics[width=0.6\linewidth]{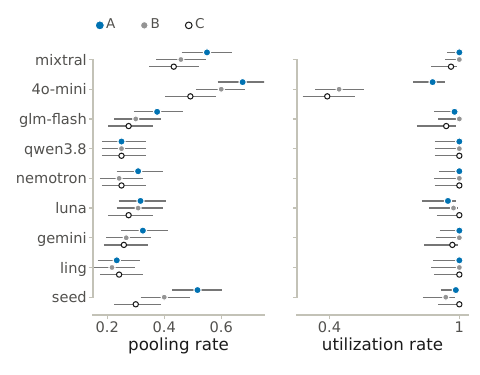}
  \caption{Evidence pooling rate (left) and evidence utilization rate (right) on the hidden profile task, one row per model in the order of Fig.~\ref{fig:F3}e, with the three instructions on each row (A blue, B grey, C open). Intervals are 95\% Wilson intervals.}
  \label{fig:B4}
\end{figure}

\subsection{Additional analyses}\label{sec:C.1}

\begin{figure}[t]
  \centering
  \includegraphics[width=\linewidth]{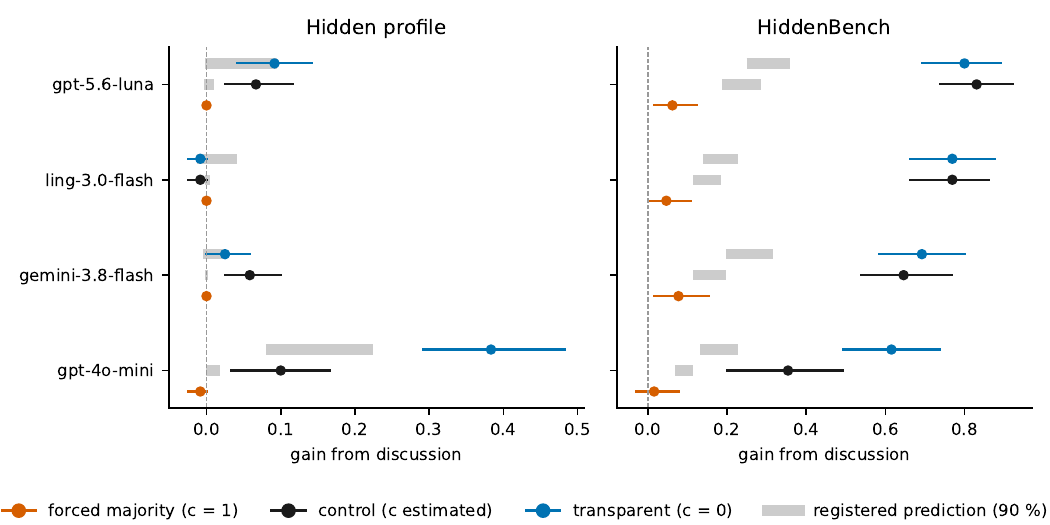}
  \caption{Gain from discussion for transparent, control, and forced-majority settings, for four LLMs on the hidden profile task and HiddenBench. Points and intervals are observations, and the gray bands are the predictions registered before the experiment.}
  \label{fig:B5}
\end{figure}

We support the main findings with five additional analyses, beginning with a direct test of causality. In experiment E1, we fixed the withholding rate $c$ in the harness and recorded our predictions before running E1.
Here $c$ is the rate at which an agent states the visible majority against its private answer.
The gain from discussion is the accuracy of the final-round majority minus that of the round-0 majority vote.
Four LLMs (gpt-4o-mini, gpt-5.6-luna, ling-3.0-flash, gemini-3.8-flash) were tested under instruction B (temperature 0.7) on 120 hidden profile tasks and 65 HiddenBench tasks.
In the transparent condition ($c = 0$), the harness replaced the public answer of each agent with its private answer.
In the forced-majority condition ($c = 1$), the harness replaced the public answer with the visible majority and kept the agent's own answer on a tie.
In the control condition, the harness left public answers unchanged.
The gain from discussion moved in the predicted direction (Fig.~\ref{fig:B5}).

Setting $c = 1$ (forced-majority) nearly eliminated the gain from discussion, as expected. Setting $c = 0$ (transparent) increased the gain only for gpt-4o-mini, the one model where public and private answers differed in the control condition. For the other three models, public and private answers already matched, so the transparent setting had little effect, as predicted.

\begin{figure}[t]
  \centering
  \includegraphics[width=\linewidth]{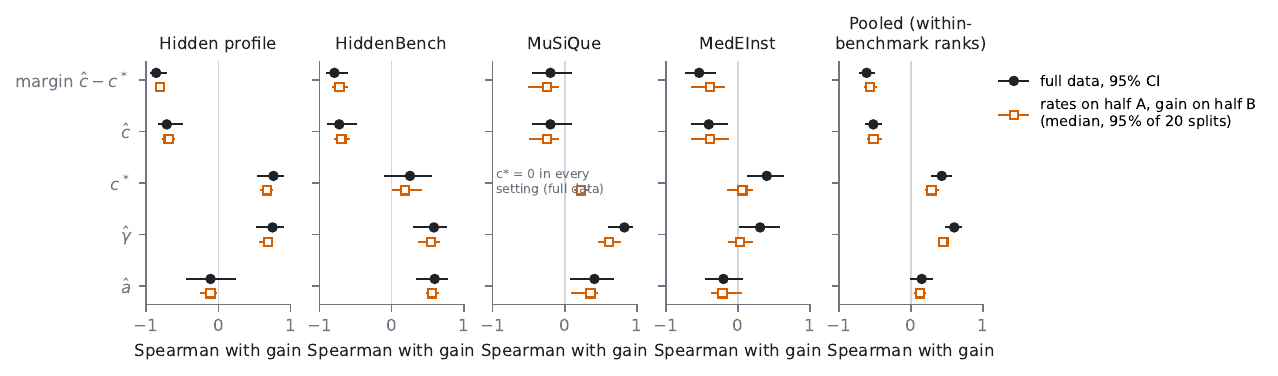}
  \caption{Rank correlations of gain with the margin and with its component rates $\hat c$, $\hat a$ and $\hat\gamma$ on four benchmarks, with a pooled row. Black circles and intervals show all-data estimates, and orange open squares and intervals show held-out task-split estimates, with medians and 95\% intervals over 20 splits.}
  \label{fig:B6}
\end{figure}

To make sure the observed ranking of settings by the margin $\hat c - c^*$ was not simply due to estimating both rates and outcomes from the same data, we ran a split-half analysis. For each benchmark, we estimated the rates on half of the tasks and measured the gain on the other half, over 20 random splits.
We also estimated the rates on rounds 1--2 and measured recovery at round 3.
The ordering of settings remained consistent across these splits.
On MuSiQue, the margin equals $\hat c$, because the net correction is negative in every setting and $c^*$ is therefore zero (Fig.~\ref{fig:B6}).
The binary model predicts no recovery when $c^* = 0$.
MuSiQue still shows a positive gain, because its free-text answers spread the incorrect answers over many strings instead of a single competitor (\appref{A.3}) and because disclosed evidence stays visible.

Because the ordering survives held-out estimation, the next question is whether the model reproduces the observed level of recovery, in addition to the ranking.
We drew parameter samples from the posterior distribution of the estimated rates ($c$, $a$ and $\gamma$ of each setting) and simulated agent-level discussion initialized with the observed round-0 answers.
We then compared the predicted and observed proportions of teams whose final majority was correct, grouped by the initial number of correct agents at round 0.

\begin{figure}[t]
  \centering
  \includegraphics[width=\linewidth]{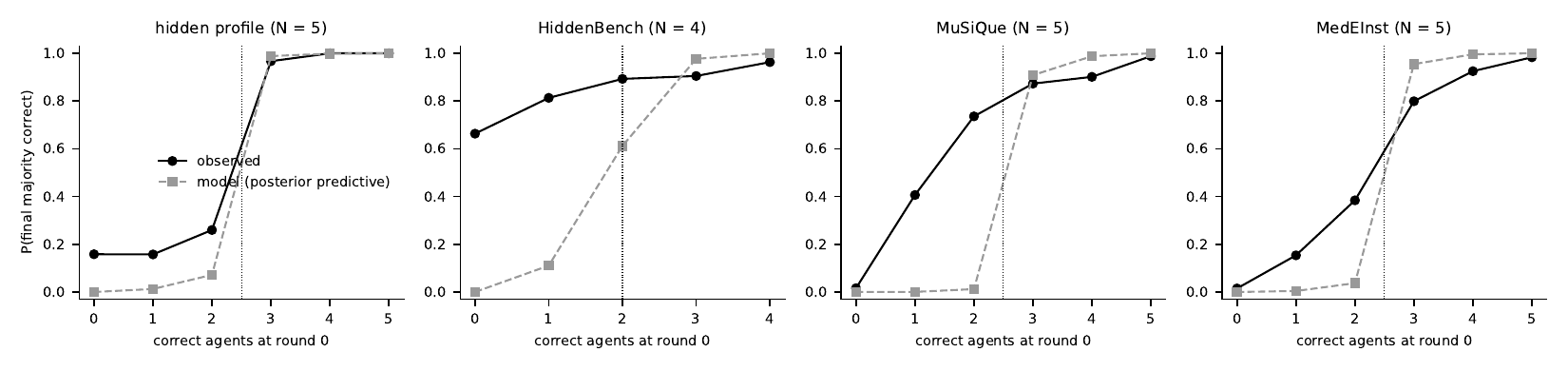}
  \caption{Probability that the final majority is correct against the number of agents correct at round 0, in panels for the hidden profile task, HiddenBench, MuSiQue, and MedEInst. Solid black marks are observations, and dashed gray marks are posterior-predictive model estimates from the model simulated with each setting's rates drawn from their posterior.}
  \label{fig:B7}
\end{figure}

The model predictions track the observations when the round-0 majority was already correct, and fall short when a team whose round-0 majority was incorrect recovers (Fig.~\ref{fig:B7}).
Such recoveries occur when evidence held privately reaches the team, and the current model includes no such recovery path.

\begin{figure}[t]
  \centering
  \includegraphics[width=\linewidth]{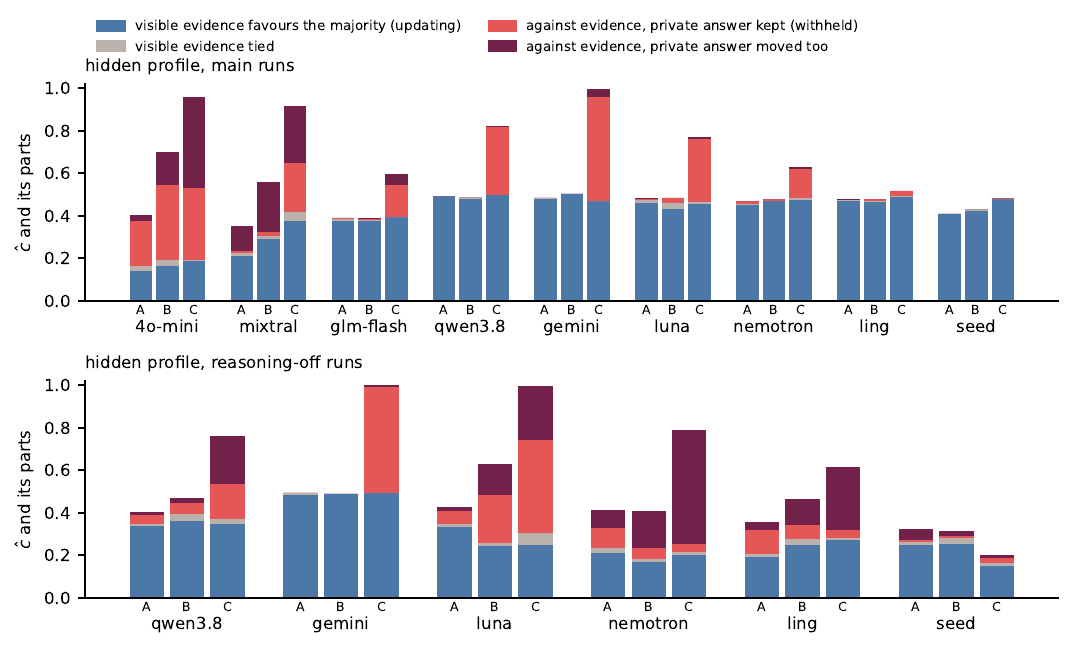}
  \caption{Stacked shares of the rounds counted as withholding on the hidden profile task. Categories separate whether visible evidence supported, tied, or opposed the majority and whether the private answer stayed opposed or moved with the majority. The upper panel shows the main runs and the lower panel the reasoning-off runs.}
  \label{fig:B8}
\end{figure}

To understand why evidence-driven recovery escapes the model, we finally decompose what the estimated withholding rate counts, the rate at which an agent whose previous private answer opposed the visible majority states the majority.
For the main runs on the hidden profile task, we partitioned the rounds counted as withholding according to whether the shared evidence visible to the agent supported the majority, was tied, or opposed it.
We further classified each round by whether the private answer moved with the majority or stayed opposed, and aggregated the shares by model and instruction.
The lower panel of Fig.~\ref{fig:B8} shows the corresponding distribution for the reasoning-off runs.

In the reasoning models the visible shared evidence and the private answer sit on the majority side of most rounds, and divergence between the two appears mainly for gpt-4o-mini and under instruction C (Fig.~\ref{fig:B8}).

\end{document}